\documentclass[aps,prl,epsfig,showpacs,floats,twocolumn,amssymb,amsmath,floatfix]{revtex4}

\usepackage{graphicx}
\usepackage{xspace}

\begin{document}

\title{Emergent Run-and-Tumble Behavior in a Simple Model of {\em Chlamydomonas} with Intrinsic Noise}

\author{Rachel R. Bennett}
\author{Ramin Golestanian}
\email[]{ramin.golestanian@physics.ox.ac.uk}
\affiliation{Rudolf Peierls Centre for Theoretical Physics, University of Oxford, Oxford OX1 3NP, UK}

\date{\today}

\begin{abstract}
Recent experiments on the green alga {\em Chlamydomonas} that swims using synchronized
beating of a pair of flagella have revealed that it exhibits a run-and-tumble behavior
similar to that of bacteria such as {\em E. Coli}. Using a simple purely hydrodynamic
model that incorporates a stroke cycle and an intrinsic Gaussian white noise, we show
that a stochastic run-and-tumble behavior could emerge, due to the nonlinearity
of the combined synchronization-rotation-translation dynamics. This suggests
the intriguing possibility that the alga might exploit nonlinear mechanics---as
opposed to sophisticated biochemical circuitry as used by bacteria---to
control its behavior.
\end{abstract}

\maketitle


Microscopic organisms need to develop swimming strategies that can tackle the low Reynolds number conditions
where viscous forces dominate \cite{lauga-powers}. Moreover, their survival crucially depends on developing
efficient search strategies for the needed chemicals \cite{search}, as has been studied extensively for many
bacteria and in particular {\em E. Coli} \cite{berg1}, which exhibits a run-and-tumble behavior \cite{run-and-tumble,R&T2}.
The pattern of having relatively long {\em run} segments in the trajectory intercalated with burstlike
{\em tumble} events during which the orientation of the swimming micro-organism is completely randomized is known to
have significant advantages as a search strategy \cite{levy}. In bacteria, the onset of the tumble events is believed
to be triggered when certain receptors report a change in the concentration of specific chemicals, which they measure
through temporal integration, via signaling pathways that have a feedback control on the preferred direction of the
rotation of the flagellar motors \cite{biochemical-circuitry}.

Recent experiments on {\em Chlamydomonas}, which is a unicellular green alga that swims with two (flexible)
flagella which beat with a breaststroke-like motion \cite{ringo:1967}, have revealed a similar pattern
of behavior \cite{ptdgg:2009}. It is observed that when the two flagella are synchronized the cell swims in
a straight line, and that the intervals of synchronous swimming are interrupted by periods of asynchronous
beating that leads to reorientation of the cell. The synchronous and asynchronous periods of beating are
analogous to the run-and-tumble motion observed in flagellated bacteria. This suggests that synchronization
of the beating flagella---which has been studied using a variety of different theoretical and experimental approaches
over the last decade \cite{ptdgg:2009,sync,ug:2011,fj:2012}---and its interplay with the swimming motion
of {\em Chlamydomonas} play a central role in controlling this stochastic behavior. A key question naturally
arises: to what extent is a direct biochemical switch necessary for the observed run-and-tumble behavior in
{\em Chlamydomonas}? In other words, is it possible to have an alternative mechanical switch that could
lead to the same behavior? This is the question we would like to address in this Letter.

\begin{figure*}[t]
\includegraphics[width=0.6\columnwidth]{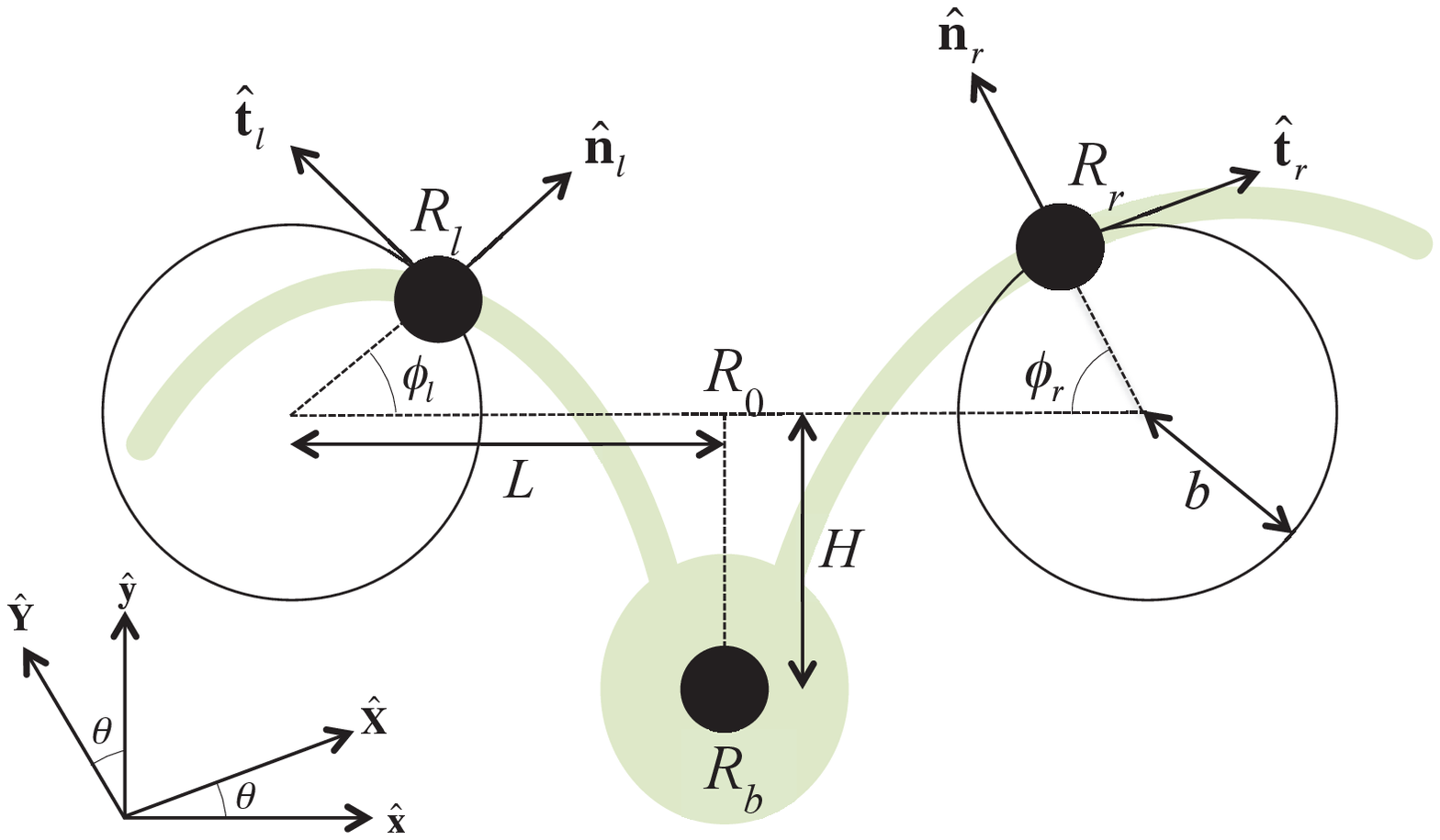}
\hspace{8.0 mm}
\includegraphics[width=0.55\columnwidth]{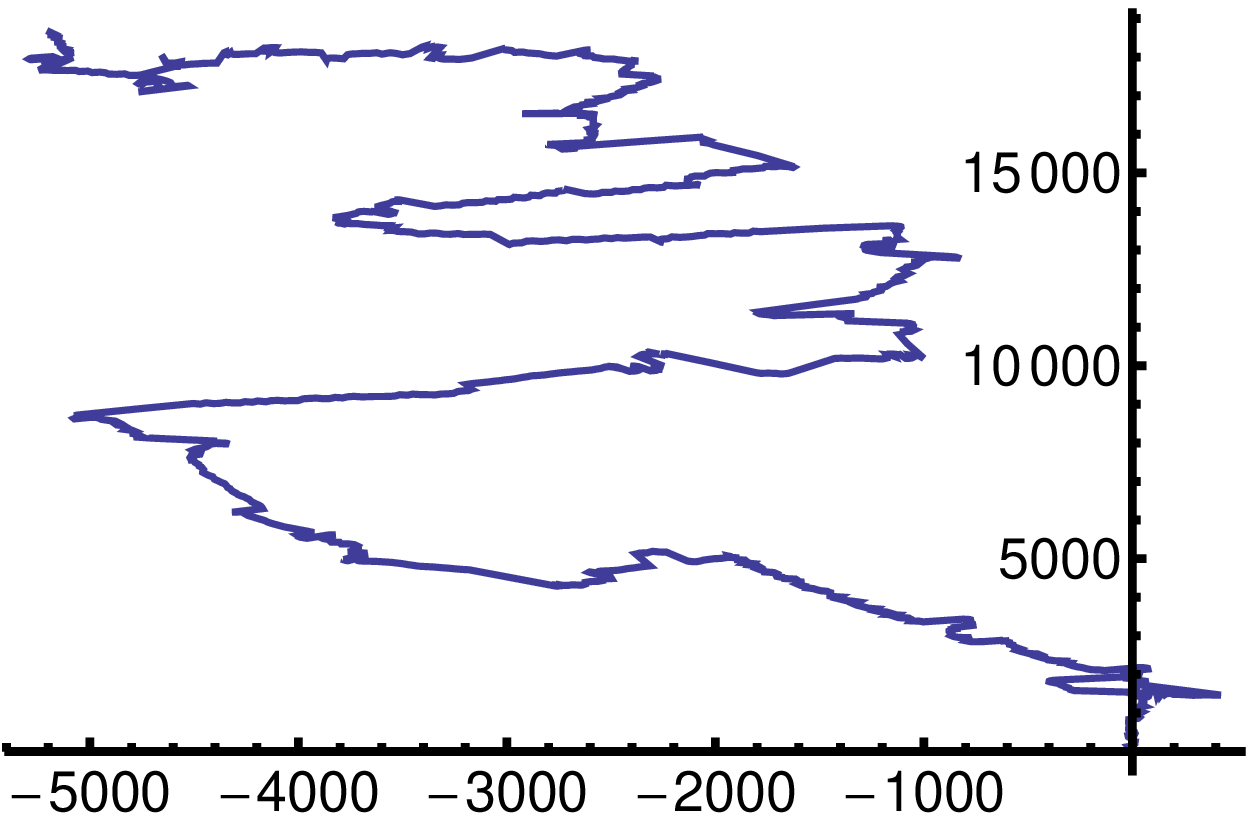}
\hspace{8.0 mm}
\includegraphics[width=0.55\columnwidth]{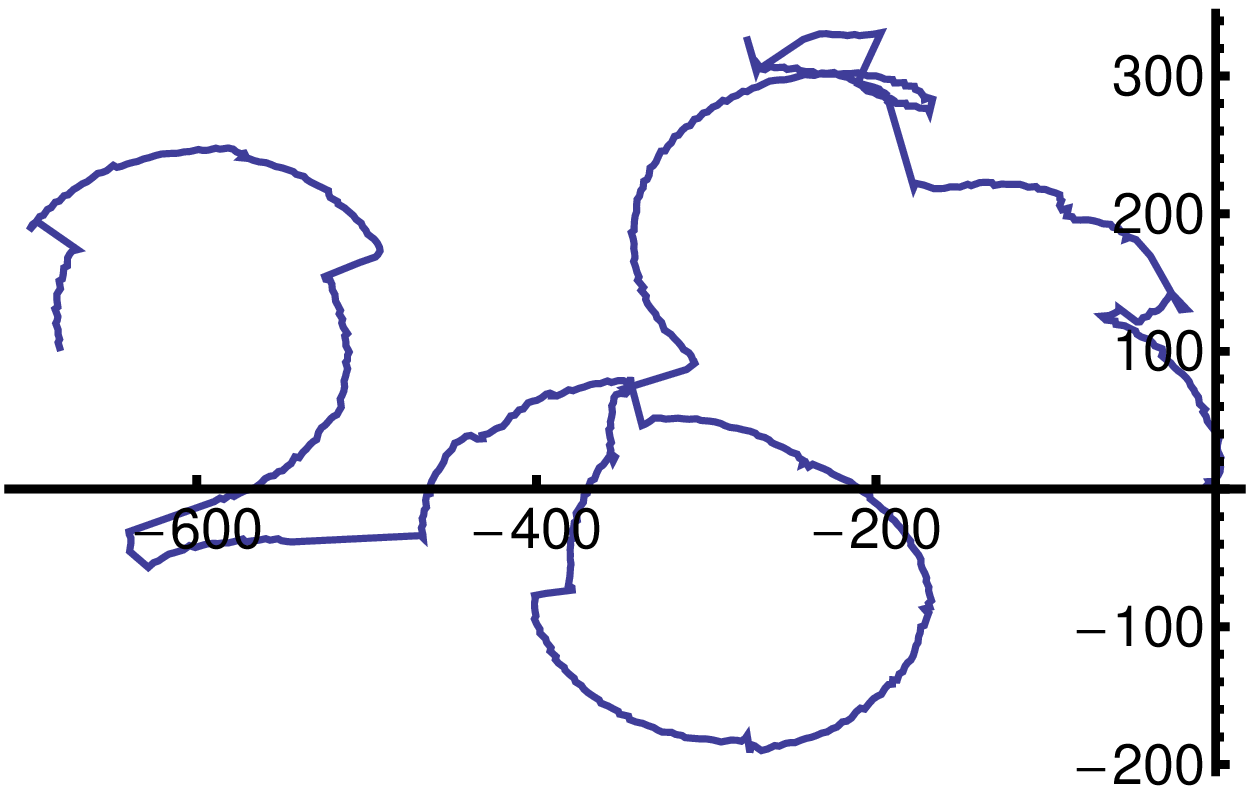} 
\\
\vspace{8.0 mm}
\includegraphics[width=0.55\columnwidth]{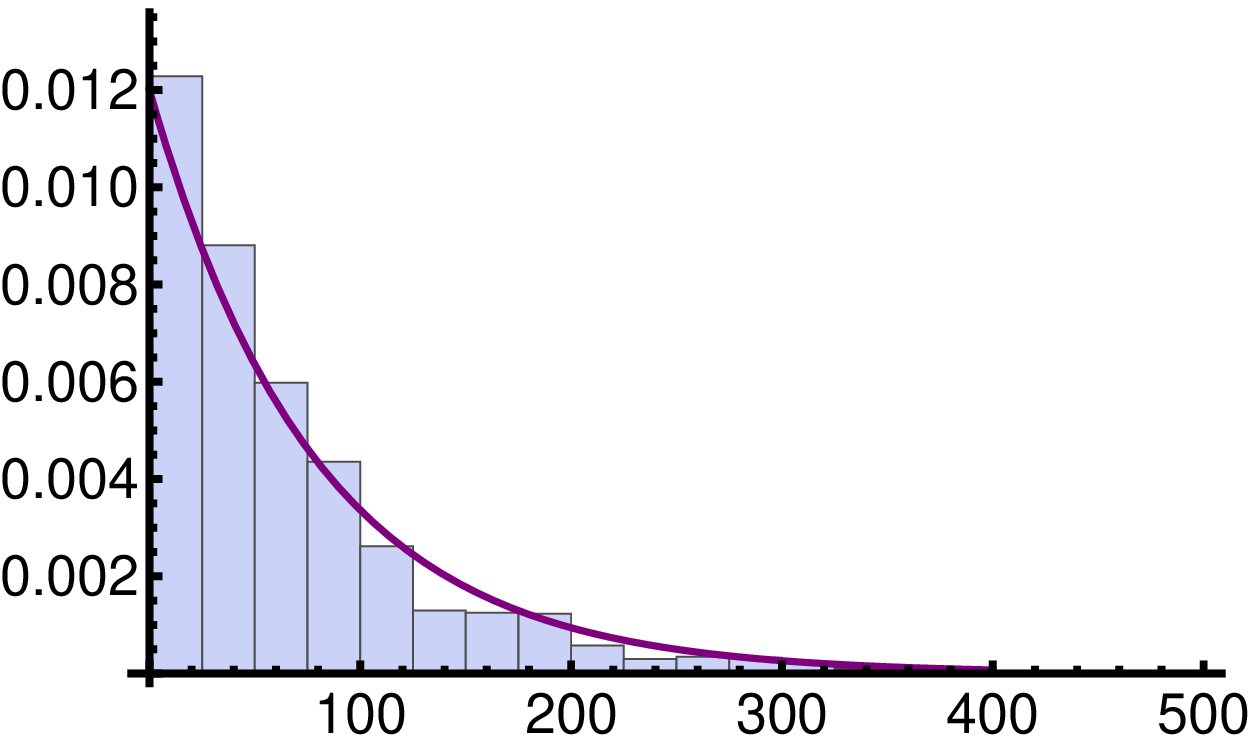}
\hspace{8.0 mm}
\includegraphics[width=0.55\columnwidth]{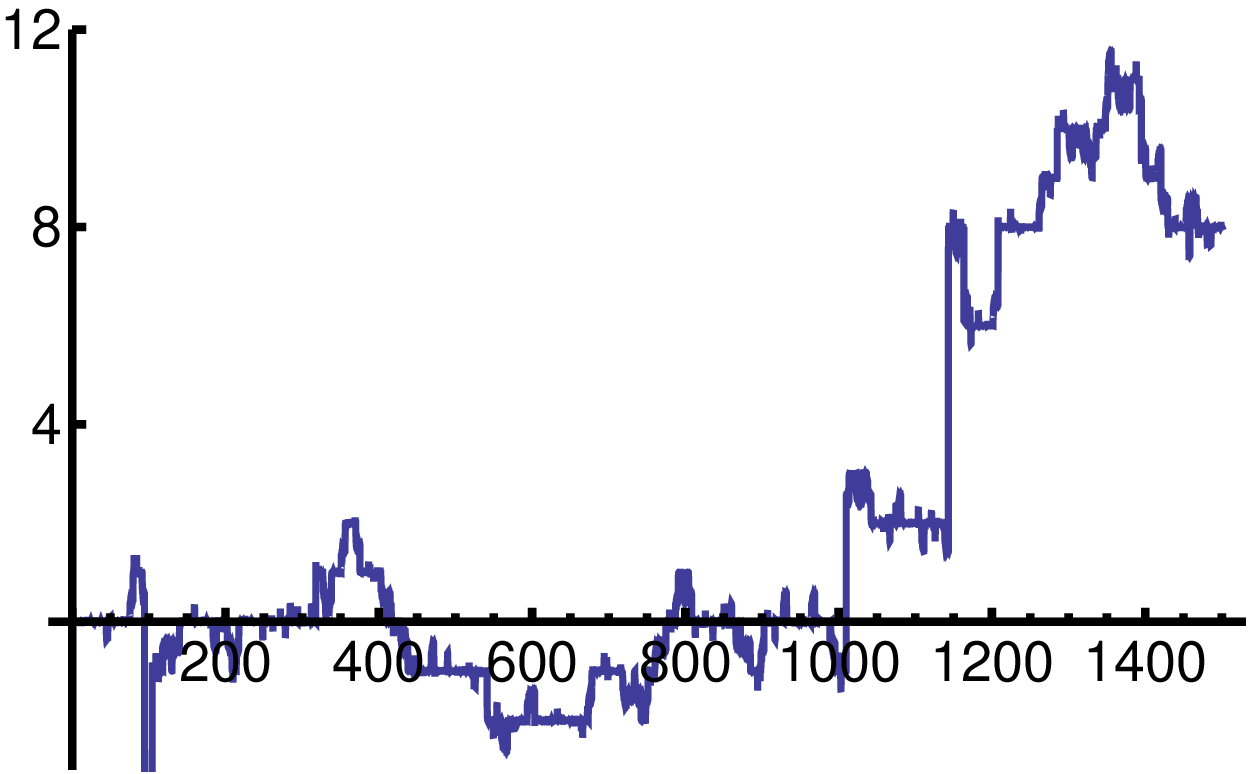}
\hspace{10.0 mm}
\includegraphics[width=0.55\columnwidth]{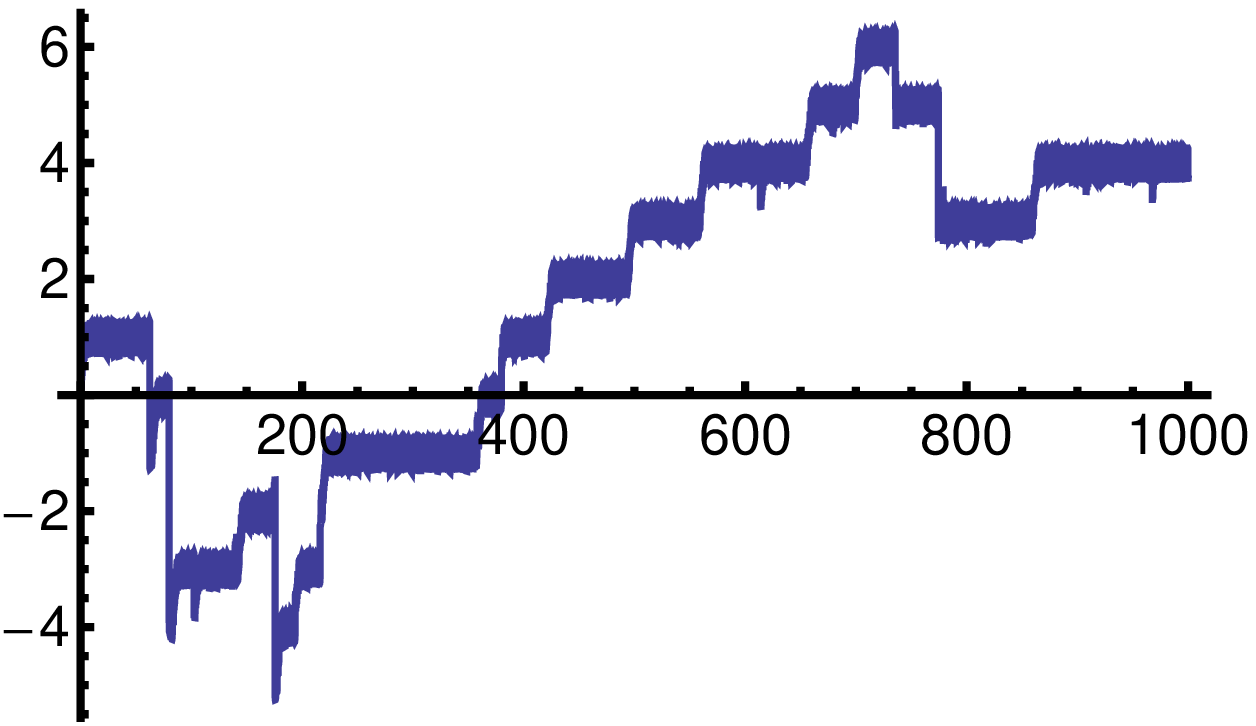} 
\\
\vspace{8.0 mm}
\includegraphics[width=0.55\columnwidth]{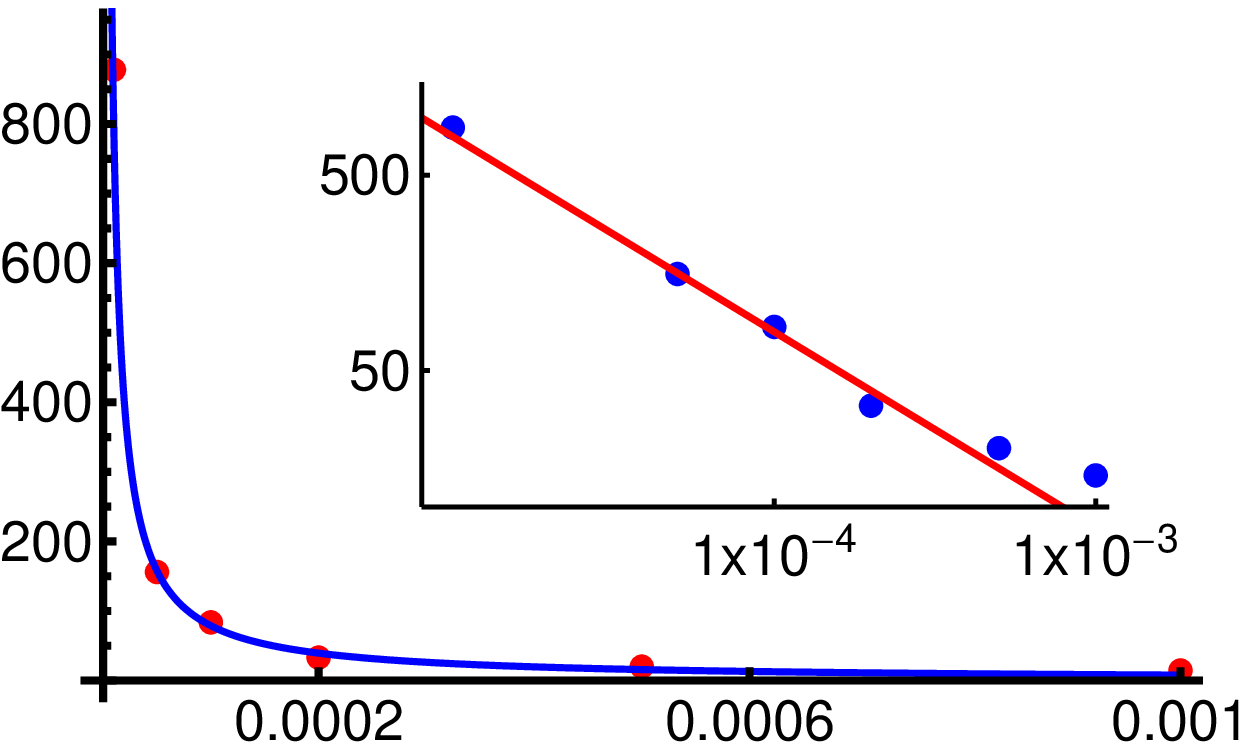}
\hspace{8.0 mm}
\includegraphics[width=0.55\columnwidth]{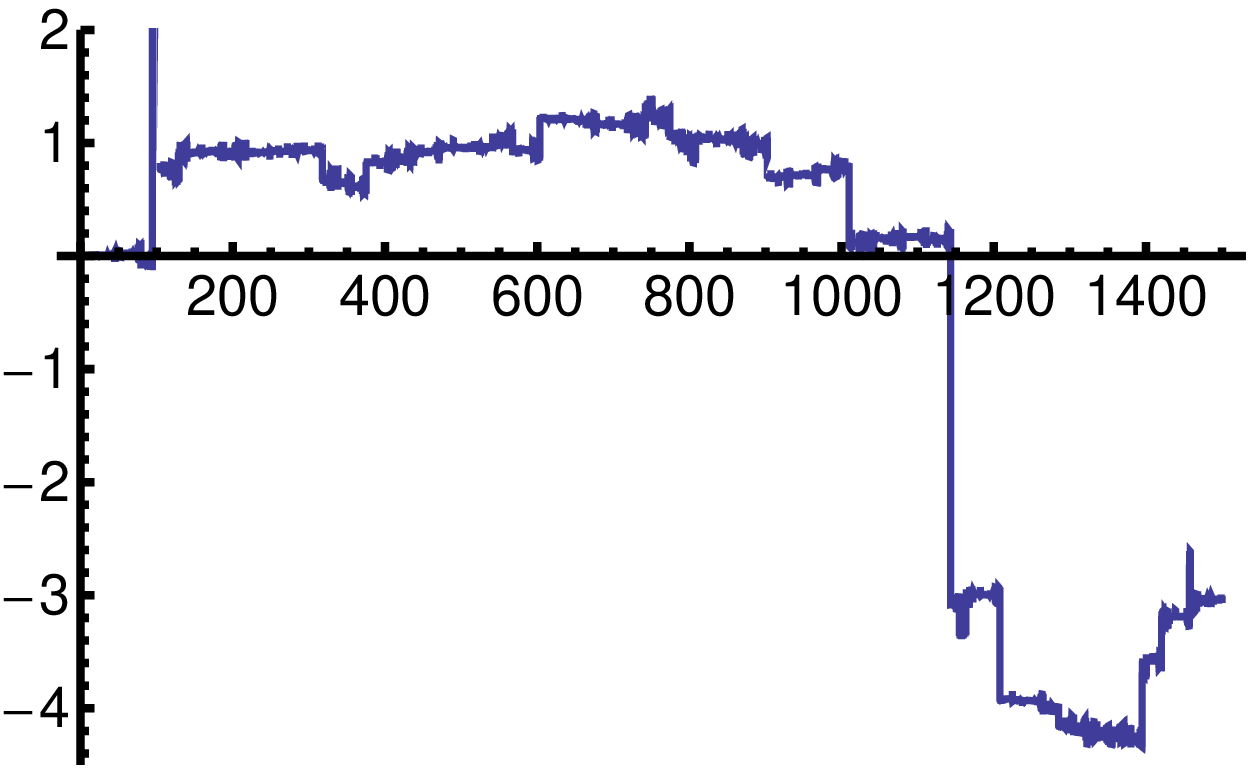}
\hspace{10.0 mm}
\includegraphics[width=0.55\columnwidth]{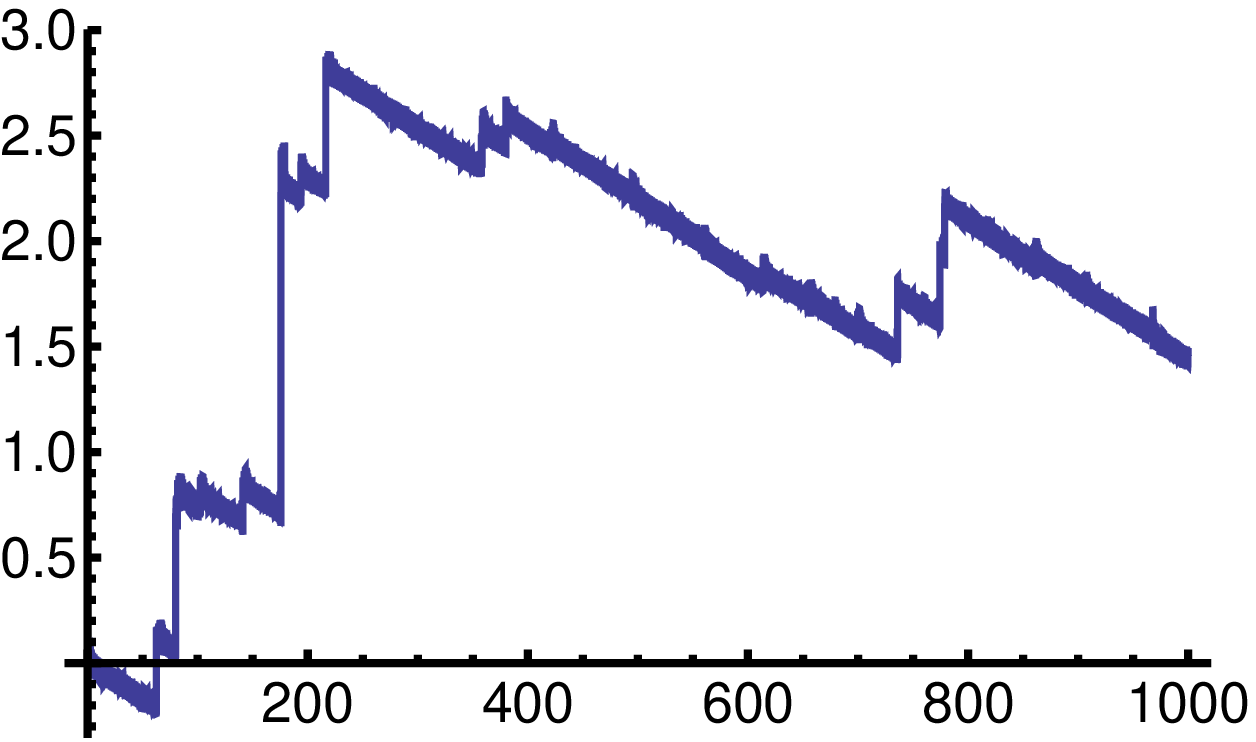} 

\begin{picture}(10,10)(0,0)

\put(-256,230){\Large{(a)}}
\put(-84,230){\Large{(d)}}
\put(98,230){\Large{(g)}}
\put(-256,124){\Large{(b)}}
\put(-84,124){\Large{(e)}}
\put(84,124){\Large{(h)}}
\put(-256,14){\Large{(c)}}
\put(-84,14){\Large{(f)}}
\put(84,14){\Large{(i)}}

\put(-240,200){\large{$P(\tau)$}}
\put(-100,108){\large{$\tau$}}

\put(-240,90){\large{\large{$\tau_{\rm run}$}}}
\put(-100,0){\large{$\sigma$}}

\put(-63,241){\large{$\frac{x}{L}$}}
\put(75,303){\large{$\frac{y}{L}$}}

\put(-76,190){\large{$\frac{\delta}{2\pi}$}}
\put(74,131){\large{$t$}}

\put(-76,84){\large{$\frac{\theta}{2\pi}$}}
\put(74,64){\large{$t$}}

\put(100,264){\large{$\frac{x}{L}$}}
\put(247,303){\large{$\frac{y}{L}$}}

\put(90,190){\large{$\frac{\delta}{2\pi}$}}
\put(244,150){\large{$t$}}

\put(90,84){\large{$\frac{\theta}{2\pi}$}}
\put(244,15){\large{$t$}}

\end{picture}
\caption{(color online.) (a) A three-sphere model of {\em Chlamydomonas}. The left and right beads represent the flagella and move on the circular trajectories shown in the cell frame. The back bead represents the cell body. The faint underlay is a schematic of a \textit{Chlamydomonas} cell. (b) Run duration statistics, in units of cycles, for driving force coefficients $a_l=a_r=0.7$ and noise strength $\sigma=10^{-4}$. The run length distribution decays as $\exp{(-\tau/\tau_{\rm run})}$ with $\tau_{\rm run} =79$ cycles in this case. (c) Dependence of $\tau_{\rm run}$ on noise, for coefficients $a_l=a_r=0.7$. $\tau_{\rm run} \propto 1/\sigma$. The inset shows the fit to a line with slope $-1$ in log-log scale. (d) A run-and-tumble trajectory for $a_l=a_r=0.7$ and $\sigma=10^{-4}$. (e) $\delta$ versus time corresponding to the trajectory in (d).
(f) $\theta$ versus time corresponding to the trajectory in (d). The jumps in $\theta$ occur at the same times as the jumps in $\delta$. (g) A run-and-tumble trajectory with curved run sections for driving force coefficients $a_l=-0.7, a_r=0.7$ and noise strength $\sigma=0.01$. (h) $\delta$ corresponding to the trajectory in (g). (i) $\theta$ corresponding to the trajectory in (g).} \label{fig:overview}
\end{figure*}

We consider a minimal three-sphere model of {\em Chlamydomonas} in which one bead models the cell body and each flagellum is represented by a bead moving on a trajectory fixed relative to the cell body, as shown in Fig. \ref{fig:overview}(a). The model
was inspired by the recent experimental measurements of the flow field around {\em Chlamydomonas} \cite{flow} that was found to
be represented reasonably well by three force monopoles (or Stokeslets) even in the near-field regions. Instead of prescribing
the motion of the beads as in Ref. \cite{ng:2004}, we treat the phases that define the locations of the flagellar beads along their trajectories as dynamical variables as in Ref. \cite{ug:2011}, and study the combined dynamics that describes their synchronization (or lack thereof) as well as rotational and translational motion of the model swimmer. Friedrich and J\"{u}licher, who independently
developed the same model in a recent paper \cite{fj:2012}, showed that the two flagella can synchronize via the coupling with
the background hydrodynamic flow, and that hydrodynamic interaction between the flagella, though essential for the net swimming,
has a secondary role in synchronization. Here, we show that by introducing a simple stroke pattern in the cyclic motion of the flagellar beads \cite{ug:2011} (as opposed to the constant forcing considered in Ref. \cite{fj:2012}) and (a relatively small amount of) Gaussian white noise, a run-and-tumble behavior emerges from the nonlinear dynamics of the model. We observe long run segments in which the model {\em Chlamydomonas} swims on straight or slightly curved trajectories (depending on the choice of parameters; see below), and intermittent sharp tumble events in which its orientation is suddenly randomized.
For a given noise strength, the distribution of the run duration decays exponentially as shown in Fig. \ref{fig:overview}(b),
in agreement with the observations by Polin {\em et al.} \cite{ptdgg:2009}. The characteristic run time $\tau_{\rm run}$
depends on the noise strength, as shown in Fig. \ref{fig:overview}(c), and follows $\tau_{\rm run} \propto 1/\sigma$,
where $\sigma$ is the noise strength. Figure \ref{fig:overview}(d) shows a typical run-and-tumble trajectory for completely symmetric driving force (explained below), which leads to run segments in which the two flagella are synchronized [Fig. \ref{fig:overview}(e)] and orientation of the swimmer is fixed [Fig. \ref{fig:overview}(f)], with large simultaneous jumps
in the relative phase and the orientation corresponding to the intermittent tumble events. A choice of driving force with
different coefficients (see below) could generate curved run segments [Fig. \ref{fig:overview}(g)] via a sequence
of phase slips [Fig. \ref{fig:overview}(h)], and in conjunction with a similar pattern of simultaneous jumps in the phase
difference [Fig. \ref{fig:overview}(h)] and the orientation [Fig. \ref{fig:overview}(i)], in agreement with the observations
of Polin {\em et al.} \cite{ptdgg:2009}.

\emph{The Model.}---Consider 3 beads in the $x-y$ plane labeled `back', `left', and `right', which we refer to with the subscripts `$b$', `$l$', and `$r$', respectively. Each bead is of radius $a$ and they are arranged in the configuration shown in Fig. \ref{fig:overview}(a). Let $\mathbf{R}_0$ be the origin of the cell frame with respect to a lab frame. The cell axes $\hat{\textbf{x}},\hat{\textbf{y}}$ make an angle $\theta(t)$ with the lab axes $\hat{\textbf{X}},\hat{\textbf{Y}}$. In the cell frame the back bead is fixed; the left and right beads that model the flagella move on circular trajectories with radius $b$, in opposite directions and with phases $\phi_l$ and $\phi_r$, respectively. The velocities of the beads are
\begin{eqnarray}
\mathbf{\dot{R}}_l&=&\mathbf{\dot{R}}_0+L\dot{\theta}\;\mathbf{\hat{y}}+b(\dot{\phi}_l-\dot{\theta})\;\mathbf{\hat{t}}_l, \label{eq:Rl} \\
\mathbf{\dot{R}}_r&=&\mathbf{\dot{R}}_0-L\dot{\theta}\;\mathbf{\hat{y}}+b(\dot{\phi}_r+\dot{\theta})\;\mathbf{\hat{t}}_r, \label{eq:Rr} \\
\mathbf{\dot{R}}_b&=&\mathbf{\dot{R}}_0-H\dot{\theta}\;\mathbf{\hat{x}}, \label{eq:Rb}
\end{eqnarray}
where the dot denotes differentiation with respect to time, and $\mathbf{\hat{n}}_i$ and $\mathbf{\hat{t}}_i$ are unit vectors in the normal and tangential directions of the circular trajectory of $\mathbf{R}_i$. The left and right beads are driven by tangential forces $F_l^t$ and $F_r^t$, respectively, which define the stroke pattern of the cyclic beatings. Normal forces $F_l^n$ and $F_r^n$ are exerted on the beads in order to constrain them to the circular trajectories. The force on the back bead is such that the swimmer is force free and torque free: $\mathbf{F}_l+\mathbf{F}_r+\mathbf{F}_b=0$, $\mathbf{T}_l+\mathbf{T}_r+\mathbf{T}_b=0$, where $\mathbf{F}_i=F_i^t\mathbf{\hat{t}}_i+F_i^n\mathbf{\hat{n}}_i$ for $i=l,r$ and $\mathbf{T}_j=\mathbf{R}_j\times \mathbf{F}_j$ for $j=b,l,r$. The forces and velocities are related through hydrodynamic interactions between the beads as
\begin{eqnarray}
\mathbf{\dot{R}}_l&=&\frac{1}{\xi}\mathbf{F}_l+(\mathbf{G}_{lr}-\mathbf{G}_{lb})\cdot \mathbf{F}_r-\mathbf{G}_{lb}\cdot \mathbf{F}_l, \label{eq:Rld} \\
\mathbf{\dot{R}}_r&=&\frac{1}{\xi}\mathbf{F}_r+(\mathbf{G}_{rl}-\mathbf{G}_{rb})\cdot \mathbf{F}_l-\mathbf{G}_{rb}\cdot \mathbf{F}_r, \label{eq:Rrd} \\
\mathbf{\dot{R}}_b&=&-\frac{1}{\xi}(\mathbf{F}_l+\mathbf{F}_r)+\mathbf{G}_{bl}\cdot \mathbf{F}_l+\mathbf{G}_{br}\cdot \mathbf{F}_r, \label{eq:Rbd}
\end{eqnarray}
where $\xi=6 \pi \eta a$ is the bead friction coefficient (with $\eta$ being the viscosity of the ambient fluid). In the limit
when $a$ is considerably smaller than the other length scales, the hydrodynamic interaction is described by the Oseen tensor
\begin{equation}
\mathbf{G}_{ij}=\frac{1}{8 \pi \eta |\mathbf{r}_{ij}|}(\mathbf{I}+\hat{\mathbf{r}}_{ij}\hat{\mathbf{r}}_{ij}),
\end{equation}
with $\mathbf{r}_{ij}=\mathbf{r}_i-\mathbf{r}_j$ \cite{Oseen}.

\emph{The Symmetric Case.}---First we consider the case where the flagella beat synchronously, and thus $\phi_l=\phi_r=\phi$. We require $F_l^t=F_r^t=F$ where $F=F(\phi)$ can vary periodically with phase and $F(\phi)>0$ for all values of $\phi$. By symmetry, $F_l^n(\phi)=F_r^n(\phi)=C(\phi)$, $\dot{\theta}=0$ and $\mathbf{\dot{R}}_b=\dot{R}_b\mathbf{\hat{y}}$. First we calculate the constraining force $C(\phi)$, and then solve for $\dot{\phi}$, which ultimately gives us the velocity $\dot{R}_b$. We find
that the cell swims with rapid oscillations with a net drift in a direction determined by the ratio $H/L$. We consider
the stability of this state in the next section. This matches the experimental observations made by Racey {\em et al.} \cite{rhn:1981}. The average velocity depends on $F(\phi)$ and the ratios $a/L$, $H/b$ and $L/b$. Here, we work with ratios $a/L=1/33$, $H/b=1/5$ and $L/b=2$. In a real {\em Chlamydomonas} cell, $a$ and $L$ are of the same order of magnitude, but
we choose small $a/L$ and $a/b$ in our calculations such that the Oseen tensor approximation is valid. The velocity scale is set by $F_0/(6 \pi \eta a)$ where $F_0=\int_0^{2\pi} d\phi F(\phi)/2\pi$; the kinematics, hydrodynamics, and the driving force profile give many additional terms to the velocity. The maximum average velocity is achieved if the driving force is constant $F(\phi)=F_0$.

\emph{Synchronization and Stability.}---We introduce a phase difference $\delta=\phi_r-\phi_l$; therefore, $\phi_l=\phi-\delta/2$ and $\phi_r=\phi+\delta/2$. The evolution of $\delta$ is governed by
\begin{eqnarray}
\dot{\delta} &=& \frac{1}{b} \Bigg\{(\dot{\mathbf{R}}_r-\dot{\mathbf{R}}_b)\cdot \hat{\mathbf{t}}_r-(\dot{\mathbf{R}}_l-\dot{\mathbf{R}}_b)\cdot \hat{\mathbf{t}}_l \nonumber \\
&&+2 \dot{\theta} \Big[\cos{(\delta/2)}\big(L\cos{\phi}-H\sin{\phi}\big)-b\Big] \Bigg\},
   \label{eq:deltadot}
\end{eqnarray}
to the leading order. We can solve Eq. (\ref{eq:deltadot}) numerically for a choice of stroke pattern (driving force) $F_i^t(\phi)$, $i=l,r$ and initial condition $\delta(\phi_0)=\delta_0$. Here, we study force profiles of the form $F_i^t(\phi)=F_0(1+a_i\cos{\phi})$ where $-1<a_i<1$.

\begin{figure}
   \begin{center}
   \includegraphics[width=0.7\columnwidth]{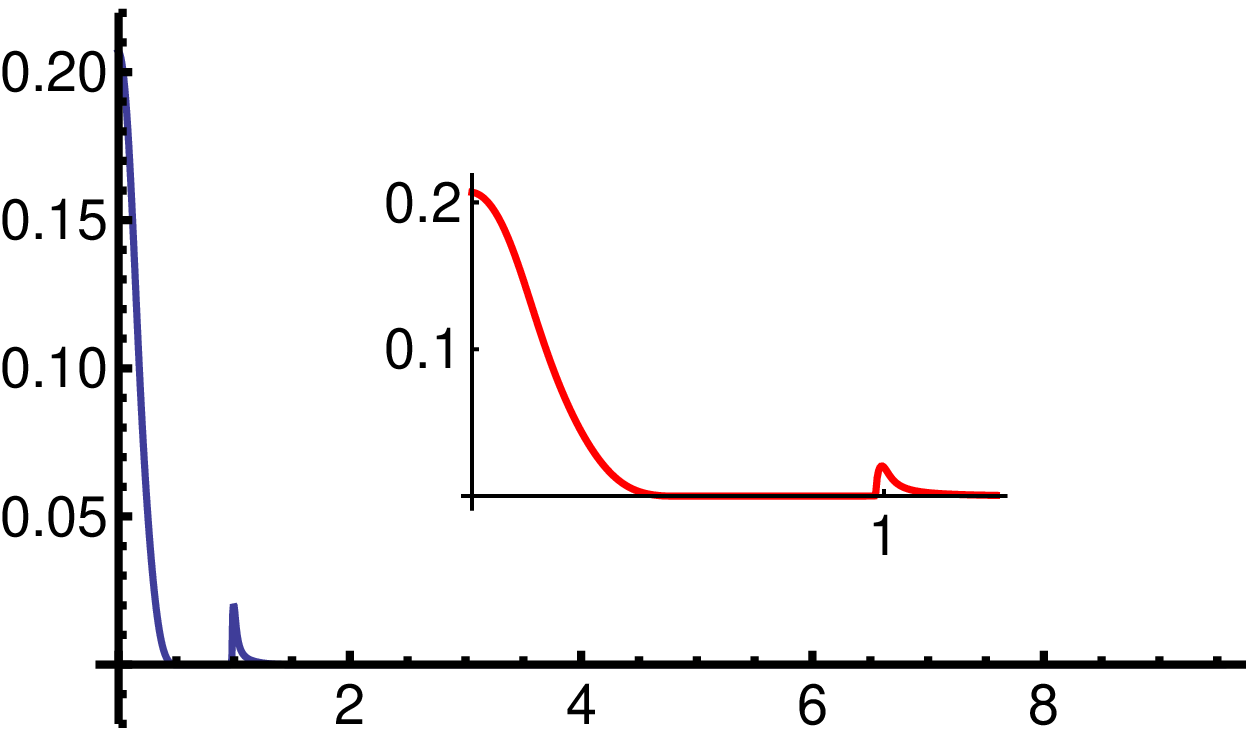}
   \\
   \vspace{4.0mm}
   \includegraphics[width=0.7\columnwidth]{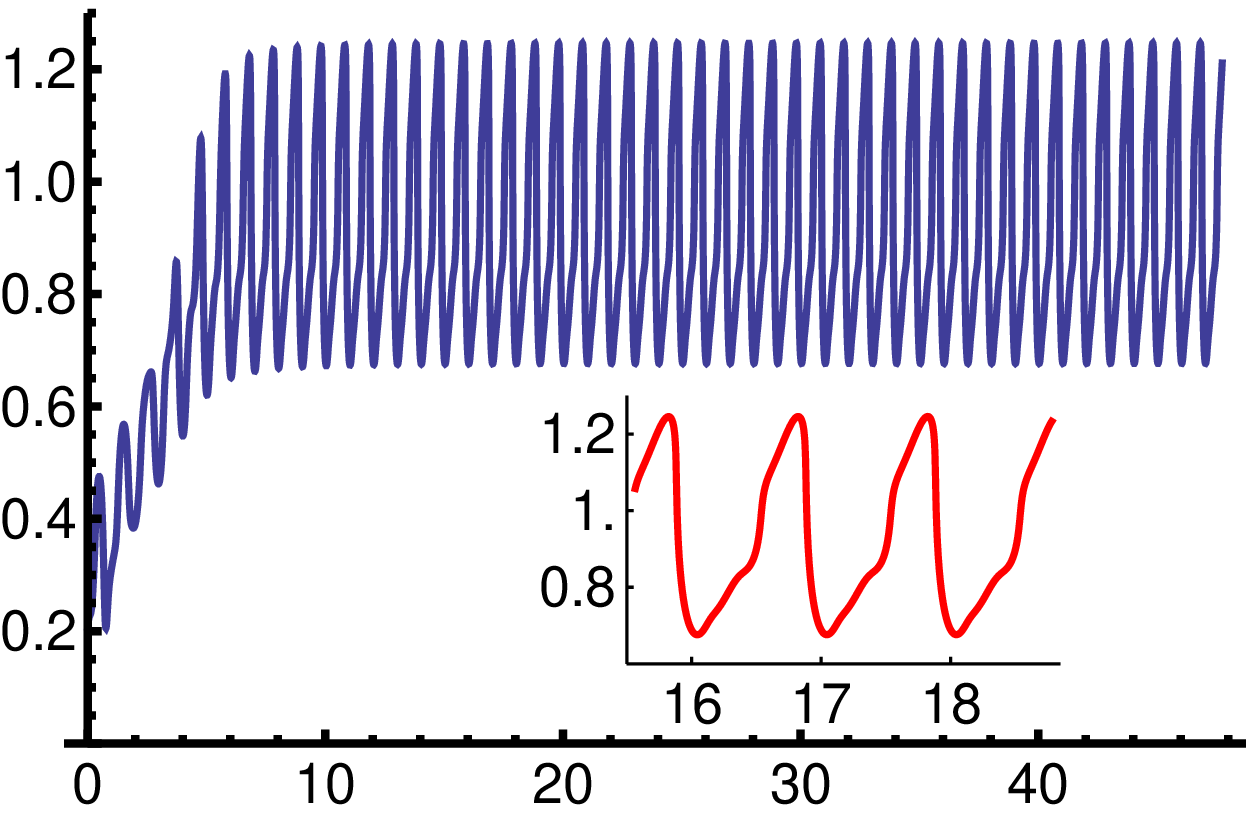}
   \\
   \vspace{6.0mm}
   \includegraphics[width=0.7\columnwidth]{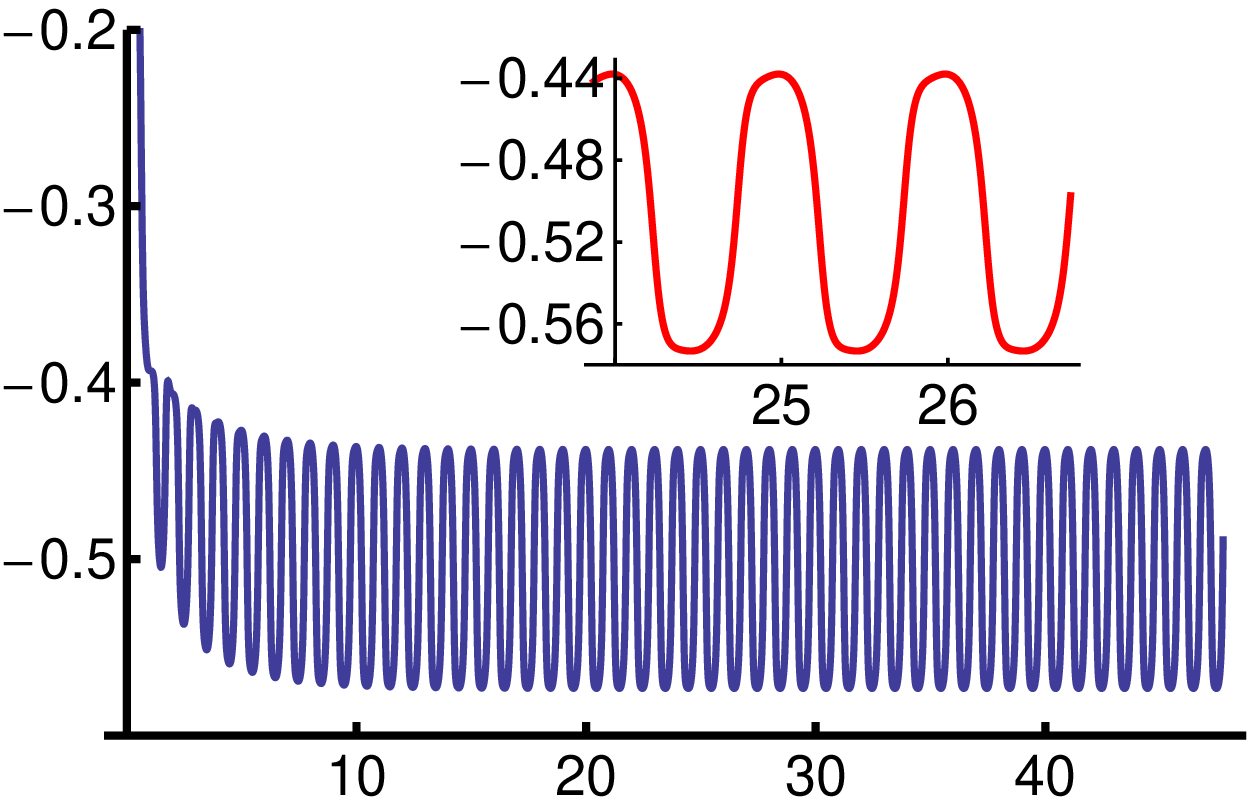}
   \end{center}
   \begin{picture}(10,10)(0,0)
   \put(93,26){\large{$\frac{\phi}{2\pi}$}}
   \put(94,156){\large{$\frac{\phi}{2\pi}$}}
   \put(94,280){\large{$\frac{\phi}{2\pi}$}}
   \put(72,78){\large{$\frac{\phi}{2\pi}$}}
   \put(68,167){\large{$\frac{\phi}{2\pi}$}}
   \put(62,303){\large{$\frac{\phi}{2\pi}$}}

   \put(-96,126){\large{$\frac{\delta}{2 \pi}$}}
   \put(-31,113){\large{$\frac{\delta}{2 \pi}$}}
   \put(-96,253){\large{$\frac{\delta}{2 \pi}$}}
   \put(-22,195){\large{$\frac{\delta}{2 \pi}$}}
   \put(-96,366){\large{$\frac{\delta}{2 \pi}$}}
   \put(-41,342){\large{$\frac{\delta}{2 \pi}$}}

   \put(-110,286){\Large{(a)}}
   \put(-110,163){\Large{(b)}}
   \put(-110,31){\Large{(c)}}
   \end{picture}
   \caption{(color online.) Evolution of $\delta$ for initial condition $\delta(\phi=0)=1.3$ with (a) $a_l=a_r=0.7$, (b) $a_l=-0.7, a_r=0.7$, (c) $a_l=0.7, a_r=0.8$. The insets show a small part of the plot in more detail. The bottom axis $\phi/2\pi$ shows the number of cycles which increases monotonically with time.}
   \label{fig:devolution}
\end{figure}

The behavior depends on the choice of coefficients $a_l$ and $a_r$. Figure \ref{fig:devolution} shows examples of the three
main types of behavior we observe. When the coefficients are equal $a_l=a_r=a_{\rm equal}$, $\delta$ evolves into the synchronized
state when $0.5 < a_{\rm equal} < 0.8$ and the initial condition is in a region around $\delta=0$.
Figure \ref{fig:devolution}(a) shows this for $a_{\rm equal}=0.7$. For other values of $a_{\rm equal}$, $\delta$ evolves into a periodically oscillating state about $\pi$. The synchronized state is stable for $0.5 < a_{\rm equal} < 0.8$ and unstable for other values of $a_{\rm equal}$. We can consider the stability of other driving force profiles with equal coefficients by studying how $\delta$ evolves for initial conditions close to $\delta=0$. For example, if we replace the $\cos{\phi}$ term in the driving force with $\cos{2\phi}$, we find that the synchronized state is stable for $a_{\rm equal} \lesssim -0.3$ and unstable otherwise.

When the coefficients have opposite sign then $\delta$ evolves into a periodically oscillating state near $2 \pi n$ (where $n$ is an integer). Figure \ref{fig:devolution}(b) shows the evolution into a state oscillating about $2 \pi$ for $a_l=-0.7, a_r=0.7$. For other values of $a_l=-a_r$, the shape of the oscillation changes and the center of the oscillation drifts, but remains close to $2 \pi n$. When the coefficients have the same sign but different magnitudes, then $\delta$ evolves into a state oscillating periodically about $\pi$, as shown in Fig. \ref{fig:devolution}(c) for $a_l=0.7,a_r=0.8$ \cite{note}.

\emph{Noise.}---The existence of well defined stable dynamical states suggests that we might obtain sharp stochastic transitions between them if we take into account an intrinsic noise in the driving force. We consider the driving force as $F_i^t(\phi_i)=1+(a_i+\zeta_i)\cos{\phi_i}$ with index $i=l,r$. $\zeta_i(t)$ has a Gaussian probability distribution with zero mean and correlation function $\langle \zeta_i(t) \zeta_j(t') \rangle = \sigma^2 \delta_{ij}\delta(t-t')$. We find that various types of behavior can occur depending on the noise strength $\sigma$ and the choice of $a_l, a_r$.

We start by considering the simplest case of $a_l=a_r=0.7$, which leads to a quick and robust synchronization as seen in Fig. \ref{fig:devolution}(a). Figure \ref{fig:overview}(d) and (e) show a cycle averaged trajectory of the cell and the evolution of the phase difference in time for $\sigma=10^{-4}$, respectively. The phase difference randomly oscillates about $2\pi n$ for a time $\tau$ before slipping away and oscillating about $2 \pi m$; with $m \neq n$. The distribution of the time between stochastic slips $\tau$ is shown in Fig. \ref{fig:overview}(b) for $\sigma=10^{-4}$ and decays exponentially, indicating a Poisson behavior and matching the observations of Polin {\em et al.} \cite{ptdgg:2009}. The characteristic run length $\tau_{\rm run}$ is inversely proportional to the noise strength $\tau_{\rm run} \propto 1/\sigma$, as seen in Fig. \ref{fig:overview}(c). Note that our numerical results for larger values of $\sigma$ deviate from the $1/\sigma$ behavior, because for those values we approach the cutoff limit in which the run length is comparable to the duration of the tumble event. Figure \ref{fig:overview}(f) shows the orientation of the cell, demonstrating that the slips in $\delta$, shown directly above in \ref{fig:overview}(e), correspond to the changes in orientation.

We next consider opposite signs on the coefficients, as $a_l=-0.7$ and $a_r=0.7$, which is interesting due to a degree of inherent frustration. In the noiseless case, $\delta$ oscillates periodically about $2 \pi n$. When noise is added, the same behavior is observed as for equal coefficients: $\delta$ oscillates about $2 \pi n$ during run phases, then slips and oscillates about $2\pi m$. Figures \ref{fig:overview}(g), \ref{fig:overview}(h), and \ref{fig:overview}(i) show a trajectory with the corresponding phase difference and orientation, respectively, for $\sigma=0.01$. Figures \ref{fig:overview}(g) and \ref{fig:overview}(i) show that the run sections always curve to the left as a result of the imbalance, which also gives a bias to slips in the positive direction where the right bead performs an additional beat. The net swimming velocity is an order of magnitude lower for opposite coefficients than for equal coefficients.

We can also obtain run-and-tumble behavior for other combinations of the coefficients. For example, for the case where we introduce a mismatch in the coefficients, say as $a_l=0.7, a_r=0.8$ [Fig. \ref{fig:devolution}(c); $\delta$ oscillates about $(2n+1)\pi$], the presence of noise introduces phase slips and we obtain run-and-tumble behavior where the flagella beat out of phase during
the run segments.

\emph{Discussion.}---We obtain run-and-tumble behavior with this simple model using only mechanical considerations that govern the force-free and torque-free swimming of the model {\em Chlamydomonas}, dynamical synchronization of the two flagella, and intrinsic noise. Remarkably, the threshold-like behavior that in bacteria such as {\em E. Coli} is believed to be controlled by a sophisticated feedback mechanism controlled by biochemical signals, emerges naturally from nonlinearities of the mechanics of the system and could be triggered by uncorrelated white noise. The parameters in the driving force do not need to be tuned to achieve run-and-tumble behavior, and could lead to large variety of behaviors such as straight and curved trajectories as well as in-phase and out-of-phase synchronization. Another interesting feature of our model is that phase slips and persistent rotation in one direction could be obtained without the need for an intrinsic frequency difference between the two flagella (i.e when the average intrinsic beating frequencies of the two flagella are the same but they correspond to opposite signs of the coefficients), which could be interpreted as if they are going through their strokes with an internal $\pi$ phase difference. In other words, the asymmetry between the flagella does not necessarily need to be intrinsic and could be dynamically enforced. Finally, the same model with a slightly different choice of (mismatched) coefficients could also describe the case of anti-phase flagellar beating during the run segments. This behavior has been recently observed in a mutant of {\em Chlamydomonas} \cite{Goldstein2}.

In conclusion, we have presented a simple model for {\em Chlamydomonas} that can describe all main experimentally observed features of the flagellar dynamics and swimming behavior of the alga. This will hopefully open up a wide range of possibilities for quantitative studies of their behavior, and help shed light on possibilities to exploit mechanical effects and constraints towards biological functions.

\acknowledgements

We would like to thank Nariya Uchida for fruitful discussions and the EPSRC for financial support.


\begin{thebibliography}{10}

\bibitem{lauga-powers}
E. Lauga and T.R. Powers, Rep. Prog. Phys. {\bf 72}, 096601 (2009).

\bibitem{search}
J.G. Mitchell, The American Naturalist {\bf 160}, 727-740 (2002).

\bibitem{berg1}
H.C. Berg, {\em E. coli} in Motion (Springer-Verlag, New York,
2004).

\bibitem{run-and-tumble}
H.C. Berg and D.A. Brown, Nature {\bf 239}, 500–504 (1972);
R.M. Macnab and D.E. Koshland Jr., Proc. Natl. Acad.
Sci. USA {\bf 69}, 2509–2512 (1972);
T.L. Min, P.J. Mears, L.M. Chubiz, C.V. Rao, I. Golding, and Y.R. Chemla,
Nature Methods {\bf 6}, 831-835 (2009).

\bibitem{R&T2}
M.J. Schnitzer, Phys. Rev. E {\bf 48}, 2553–2568 (1993);
J. Tailleur and M.E. Cates, Phys. Rev. Lett. {\bf 100}, 218103 (2008);
Y. Kafri and R.A. da Silveira, Phys. Rev. Lett. {\bf 100}, 238101 (2008).

\bibitem{levy}
O. Benichou, M. Coppey, M. Moreau, P.-H. Suet, and R. Voituriez, Phys. Rev. Lett. {\bf 94}, 198101 (2005);
F. Bartumeus and S.A. Levin, Proc. Natl. Acad. Sci. USA {\bf 105}, 19072 (2008);
D.V. Nicolau Jr., J.P. Armitage, and P.K. Maini, Comput. Biol. Chem. {\bf 33}, 269 (2009);
F. Thiel, L. Schimansky-Geier, and I.M. Sokolov, Phys. Rev. E {\bf 86}, 021117 (2012).

\bibitem{biochemical-circuitry}
P. Cluzel, M. Surette, and S. Leibler, Science {\bf 287}, 1652-1655 (2000);
T.-M. Yi, Y. Huang, M.I. Simon, and J. Doyle, Proc. Natl. Acad. Sci. USA {\bf 97}, 4649-4653 (2000);
T.L. Min, P.J. Mears, I. Golding, and Y.R. Chemla, Proc. Natl. Acad. Sci. USA {\bf 109}, 9869-9874 (2012).

\bibitem{ringo:1967}
D.L. Ringo, J. Cell. Biol. {\bf 33}, 543 (1967).

\bibitem{ptdgg:2009}
M. Polin, I. Tuval, K. Drescher, J.P. Gollub and R.E. Goldstein, Science {\bf 325}, 487 (2009);
R.E. Goldstein, M. Polin and I. Tuval, Phys. Rev. Lett. {\bf 103}, 168103 (2009).


\bibitem{sync}
M.J. Kim, J.C. Bird, A.J. Van Parys, K.S. Breuer, and T.R. Powers,, Proc. Natl. Acad. Sci. USA {\bf 100}, 15481 (2003);
M.C. Lagomarsino, P. Jona, and B. Bassetti, Phys. Rev. E {\bf 68}, 021908 (2003);
M. Kim and T.R. Powers, Phys. Rev. E {\bf 69}, 061910 (2004);
M. Reichert and H. Stark, Eur. Phys. J. E {\bf 17}, 493 (2005);
A. Vilfan and F. J\"ulicher, Phys. Rev. Lett. {\bf 96}, 058102 (2006);
Y.W. Kim and R.R. Netz, Phys. Rev. Lett. {\bf 96}, 158101 (2006);
B. Guirao and J-F. Joanny, Biophys. J. {\bf 92}, 1900-1917 (2007);
T. Niedermayer, B. Eckhardt, and P. Lenz, Chaos {\bf 18}, 037128 (2008);
G. J. Elfring and E. Lauga, Phys. Rev. Lett. {\bf 103}, 088101 (2009);
N. Uchida and R. Golestanian, Phys. Rev. Lett. {\bf 104}, 178103 (2010);
J. Kotar, M. Leoni, B. Bassetti, M. C. Lagomarsino, and P. Cicuta,
Proc. Natl. Acad. Sci. USA {\bf 107}, 7669 (2010);
R. Di Leonardo, A. Buzas, L. Kelemen, G. Vizsnyiczai, L. Oroszi, and P. Ormos,
Phys. Rev. Lett. {\bf 109}, 034104 (2012).

\bibitem{ug:2011}
N. Uchida and R. Golestanian, Phys. Rev. Lett. {\bf 106}, 058104 (2011); arXiv:1209.4481v1.

\bibitem{fj:2012}
B.M. Friedrich and F. J\"{u}licher, Phys. Rev. Lett. {\bf 109}, 138102 (2012).

\bibitem{flow}
K. Drescher, R.E. Goldstein, N. Michel, M. Polin, and I. Tuval, Phys. Rev. Lett. {\bf 105}, 168101 (2010);
J.S. Guasto, K.A. Johnson, and J.P. Gollub, Phys. Rev. Lett. {\bf 105}, 168102 (2010).

\bibitem{ng:2004}
A. Najafi and R. Golestanian, Phys. Rev. E {\bf 69}, 062901 (2004).

\bibitem{Oseen}
C.W. Oseen, {\em Neuere Methoden und Ergebnisse in der Hydrodynamik}
(Akademishe Verlagsgesellschaft, Leipzig, 1927).

\bibitem{rhn:1981}
T.J. Racey, R. Hallett and B. Nickel, Biophys. J. {\bf 35}, 557 (1981).

\bibitem{note}
A more detailed analysis of the behavior of the synchronization dynamics 
as a function of the form and harmonic of the stroke pattern, as well as 
the initial conditions will be presented elsewhere.

\bibitem{Goldstein2}
R.E. Goldstein, private communication.

\end{thebibliography}
\end{document}